\pdfoutput=1
\documentclass[aps,prd,amsmath,floats,floatfix, twocolumn,
superscriptaddress,nofootinbib,showpacs,longbibliography]{revtex4-1}
\usepackage[T1]{fontenc}
\usepackage[utf8]{inputenc}
\usepackage{txfonts}
\usepackage{verbatim}
\usepackage{placeins}
\usepackage[dvipsnames, usenames]{xcolor}
\definecolor{linkcolor}{rgb}{0.0,0.3,0.5}
\usepackage[hypertexnames=false, unicode, colorlinks=true, linkcolor=linkcolor,
citecolor=linkcolor, filecolor=linkcolor,urlcolor=linkcolor,
pdfusetitle]{hyperref}
\usepackage[all]{hypcap}
\usepackage{graphicx}
\usepackage{xspace}
\usepackage{amssymb}
\usepackage{microtype}
\usepackage{subfigure}
\usepackage{url}
\usepackage[english]{babel}
\usepackage{blindtext}
\usepackage[normalem]{ulem} 
\usepackage{bm} 
\usepackage{mathrsfs}
\usepackage{xspace} 

\DeclareMathAlphabet{\mathpzc}{OT1}{pzc}{m}{it}

\usepackage[nomessages]{fp}

\begin{document}
\title{Consistency of spin effects between numerical relativity and perturbation theory \\for inspiraling comparable-mass black hole binaries}

\newcommand{\KITP}{\affiliation{Kavli Institute for Theoretical Physics, University of California Santa Barbara, Kohn Hall, Lagoon Rd, Santa Barbara, CA 93106}} 
\author{Tousif Islam}
\email{tousifislam@ucsb.edu}
\KITP

\author{Gaurav Khanna} \affiliation{Department of Physics and Institute for AI \& Computational Research, University of Rhode Island, Kingston, RI 02881}
\affiliation{Department of Physics and Center for Scientific Computing \& Data Science Research, University of Massachusetts, Dartmouth, MA 02747}

\author{Scott E. Field} \affiliation{Department of Mathematics and Center for Scientific Computing \& Data Science Research, University of Massachusetts, Dartmouth, MA 02747}

\hypersetup{pdfauthor={Islam et al.}}
\date{\today}

\begin{abstract}
Numerical relativity (NR) provides the most accurate waveforms for comparable-mass binary black holes but becomes prohibitively expensive for increasingly asymmetric mass ratios. Point-particle black hole perturbation theory (ppBHPT), which expands the Einstein equations in the small-mass-ratio limit, offers a computationally efficient alternative but is expected to break down in the comparable-mass regime because it neglects nonlinear effects. Nonetheless, several recent studies have shown that ppBHPT can model non-spinning binaries with high accuracy when supplemented by simple calibrations or a first post-adiabatic (PA) correction. Here we assess the applicability of ppBHPT to quasi-circular binaries with a non-precessing primary and a non-spinning secondary by comparing waveform amplitudes, orbital frequencies, and phases. We find that spin effects in ppBHPT waveforms (without additional spin information beyond adiabatic order) agree surprisingly well with NR (outperforming some post-Newtonian models) over the last $\approx 20$ orbital cycles. This suggests that, after incorporating higher-order non-spinning corrections into ppBHPT—via second-order self-force results or semi-analytical fits—only modest spin-dependent adjustments may be required to achieve NR-faithful waveforms. We also show that combining non-spinning NR information with adiabatic ppBHPT yields reasonably accurate inspiral waveforms for spins $\chi \lesssim 0.5$ and mass ratios $q \gtrsim 5$.
\end{abstract}

\maketitle
\noindent\textbf{\textit{Introduction}: }
Numerical relativity (NR)~\cite{Mroue:2013xna,Boyle:2019kee,Healy:2017psd,Healy:2019jyf,Healy:2020vre,Healy:2022wdn,Jani:2016wkt,Hamilton:2023qkv,Scheel:2025jct} and adiabatic, small-mass-ratio point-particle perturbation theory (ppBHPT)~\cite{Aleman:2003, Khanna:2004,Burko:2007,Sundararajan:2008zm,Sundararajan:2010sr,Zenginoglu:2011zz,Fujita:2004rb,Fujita:2005kng,Mano:1996vt,throwe2010high,OSullivan:2014ywd,Drasco:2005kz} are two distinct frameworks used to compute inspiral-merger-ringdown gravitational waveforms from binary black holes (BBHs). These approaches differ in both methodology and domain of validity. While NR solves the full Einstein equations governing the spacetime without approximation, ppBHPT solves them perturbatively to low order\footnote{Throughout this paper, our ppBHPT waveforms are computed with linear perturbation theory where the smaller black hole undergoes adiabatic inspiral driven by radiative energy and angular momentum losses.} in the mass ratio. In addition, ppBHPT models the smaller (secondary) black hole, of mass $m_2$, as a point particle with no internal structure moving in the spacetime of a larger Kerr black hole with mass $m_1$ and spin $\chi$. These approximations imply that ppBHPT is expected to break down in the comparable-mass regime and deviate from NR.
Despite this expectation, several studies have shown that for non-spinning binaries ppBHPT with corrections can generate accurate waveforms in the comparable-to-intermediate mass-ratio regime. Refs.~\cite{Pound:2021qin,Miller:2020bft,Wardell:2021fyy} incorporated second-order self-force calculations for Schwarzschild binaries, while Refs.~\cite{Islam:2022laz,Rifat:2019ltp} proposed an empirical mapping between ppBHPT and NR waveforms using time-independent fitting parameters. Ref.~\cite{vandeMeent:2020xgc} extracted next-to-leading-order orbital phase corrections directly from NR waveform data, showing higher-order effects to be small. Collectively, these results demonstrate that ppBHPT—with phenomenological fits, self-force corrections, or NR-informed phase adjustments—remains accurate in the comparable-to-intermediate mass-ratio regime over the duration typically accessible to NR.

It is not yet clear whether these results extend to a spinning primary, where waveform structure becomes more complex~\footnote{Refs.~\cite{Mathews:2021rod,Mathews:2025nyb} have incorporated second-order self-force corrections for binaries with a non-spinning primary and a spinning secondary.}. However, there are indications that ppBHPT remains effective when spin is included. For example, the empirical mapping technique~\cite{Islam:2022laz,Rifat:2019ltp} has been extended to binaries with a Kerr primary and a non-spinning secondary~\cite{Rink:2024swg}, where the fitting parameters depend only weakly on the primary spin. In parallel, Ref.~\cite{vandeMeent:2023ols} showed that incorporating non-spinning second-order self-force results into semi-analytical frameworks such as the effective-one-body (EOB) model~\cite{Buonanno:1998gg, Buonanno:2000ef} improves agreement with NR even for spinning binaries. These findings suggest that spin effects may already be reasonably captured by the adiabatic ppBHPT framework throughout the inspiral and over a wide range of spin.

In this paper, we investigate the agreement between NR and ppBHPT waveforms in terms of spin effects for binaries with a non-precessing primary and a non-spinning secondary. Our approach focuses on identifying which physical effects are included or missing in the two frameworks—particularly mass ratio ($q=m_1/m_2 \geq 1$), spin ($\chi$), and their couplings. NR captures mass-ratio dependence, spin, and nonlinear interactions in full, limited only by numerical accuracy. In contrast, ppBHPT includes spin dependence through the Teukolsky equation without restriction, but only captures leading-order mass-ratio effects and omits higher-order mass–spin interactions. We expect these higher-order effects to be subdominant and demonstrate this below.

\begin{figure}
\includegraphics[width=\columnwidth]{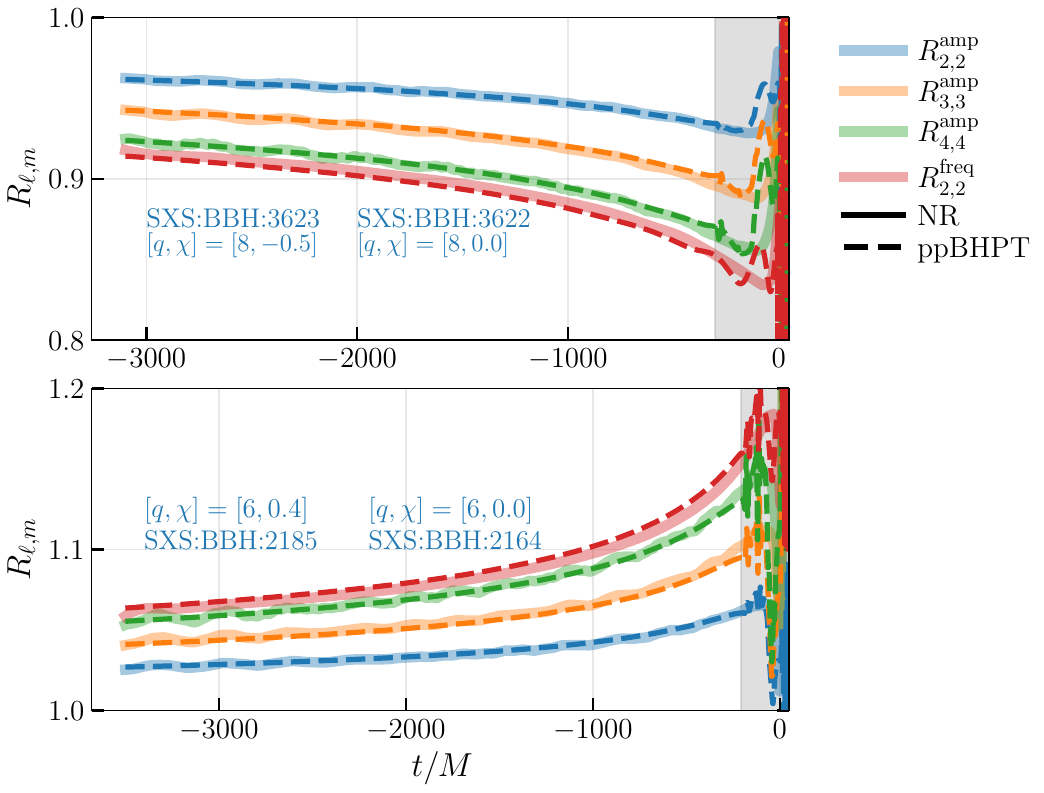}
\caption{Ratios of waveform amplitudes and frequencies defined in Eq.~\eqref{eq:sef} for two binaries: $[q,\chi]=[8,-0.5]$ (upper) and $[6,0.4]$ (lower). The quantities shown are $R_{2,2}^{\mathrm{amp}}$, $R_{3,3}^{\mathrm{amp}}$, $R_{4,4}^{\mathrm{amp}}$, and $R_{2,2}^{\mathrm{freq}}$. Solid curves denote NR results from SXS simulations, and dashed curves show adiabatic ppBHPT predictions. The adiabatic ppBHPT ratios closely track NR throughout the inspiral, with larger deviations appearing once the ppBHPT systems enter the geodesic plunge (gray regions). NR waveforms exhibit additional oscillations absent in ppBHPT, possibly due to residual eccentricity. The SXS simulation IDs used are indicated by color-coded labels.}
\label{fig:R_nr_bhpt_example}
\end{figure}

\noindent\textbf{\textit{Methodology and diagnostics}: }
We obtain NR data, simulated with the Spectral Einstein Code (\texttt{SpEC})~\cite{Kidder:2000yq}, from the publicly available SXS catalog~\cite{Scheel:2025jct,Mroue:2013xna,Boyle:2019kee}. Adiabatic ppBHPT waveforms are generated using the time-domain inspiral-merger-ringdown Teukolsky solver developed in Refs.~\cite{Khanna:2004,Burko:2007,Sundararajan:2008zm,Sundararajan:2010sr,Ori:2000zn,Hughes:2019zmt,Apte:2019txp}, with radiative energy and angular momentum losses computed using the open-source code GremlinEq~\cite{gremlin,OSullivan:2014ywd,Drasco:2005kz}. Throughout, we scale both NR and ppBHPT waveforms by the total mass $M=m_1+m_2$ and use geometric units with $G=c=1$.
To generate ppBHPT waveforms, we compute the adiabatic inspiral and smoothly attach a late-stage geodesic plunge. The transition region, which depends on mass ratio and spin, is constructed using the generalized Ori-Thorne procedure~\cite{Ori:2000zn,Hughes:2019zmt,Sundararajan:2010sr,Apte:2019txp}. Imperfections in this procedure can introduce small nonphysical oscillations near the onset of the transition in some modes, particularly for comparable-mass binaries (see Fig.~1 of Ref.~\cite{Islam:2022laz}). 

\begin{figure}
\includegraphics[width=\columnwidth]{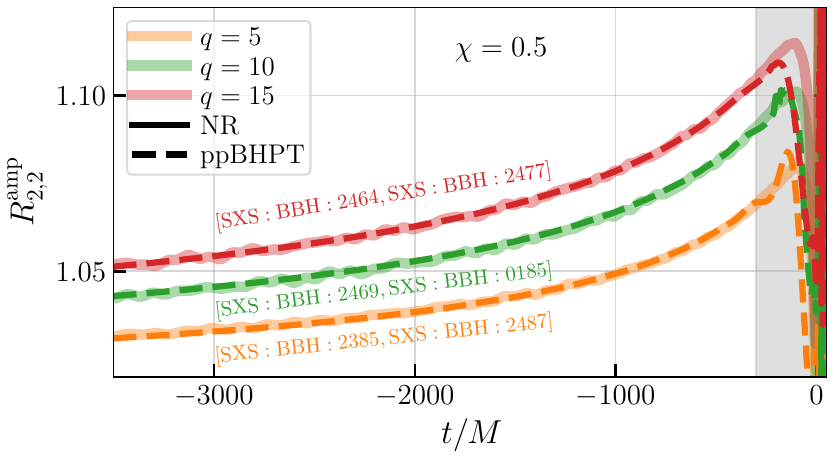}
\caption{Ratios of $(2,2)$ waveform mode amplitudes defined in Eq.~\eqref{eq:sef} for three binaries with mass ratios $q = [5, 10, 15]$ and fixed spin $\chi=0.5$. Solid curves show NR results from SXS simulations, and dashed lines show ppBHPT results. The ppBHPT ratios closely track NR during the inspiral, with larger deviations appearing once the ppBHPT systems enter the geodesic plunge (gray regions). NR waveforms exhibit additional oscillations at early and intermediate times, likely caused by residual eccentricity, which are absent in ppBHPT. We also mention the corresponding combination of spinning and non-spinning SXS NR simulation ID, as color-coded text.}
\label{fig:q_5_10_15}
\end{figure}

We write the full complex waveform as
\begin{align}
h(t, \theta, \phi; \boldsymbol{\lambda}) &= \sum_{\ell=2}^\infty \sum_{m=-\ell}^{\ell} h_{\ell m}(t; \boldsymbol\lambda)\, {}_{-2}Y_{\ell m}(\theta,\phi),
\label{hmodes}
\end{align}
where $(\ell,m)$ label spin-weighted spherical harmonic modes~\cite{Maggiore:2007ulw, Maggiore:2018sht,RevModPhys.52.299}, $t$ is time, $(\theta,\phi)$ are angles, and $\boldsymbol{\lambda}$ denotes intrinsic parameters $(q,\chi)$. We focus on binaries with a spinning primary and non-spinning secondary, and set the peak of the $(2,2)$ mode amplitude at $t=0$.
To analyze spin effects, we decompose each mode into a real-valued amplitude and phase, $h_{\ell m}(t; \boldsymbol{\lambda}) = A_{\ell m}(t)\, e^{i \phi_{\ell m}(t)}$,
with instantaneous frequency $\omega_{\ell m}(t; \boldsymbol{\lambda}) = \frac{d\phi_{\ell m}(t)}{dt}$.
We monitor the effect of spin in an NR waveform by computing the ratios 
\begin{equation} \label{eq:sef}
    R_{\ell m}^{\mathrm{NR, amp}}(t) = \frac{A_{\ell m}^{\mathrm{NR}}(t; \chi)}{A_{\ell m}^{\mathrm{NR}}(t; \chi = 0)}, \quad
    R_{\ell m}^{\mathrm{NR, freq}}(t) = \frac{\omega_{\ell m}^{\mathrm{NR}}(t; \chi)}{\omega_{\ell m}^{\mathrm{NR}}(t; \chi = 0)},
\end{equation}
of the waveform amplitude and frequency for a spinning binary relative to its non-spinning counterpart.
We refer to these ratios as \textit{spin-enhancement factors}. Similarly, we define corresponding spin-enhancement factors, $R_{\ell m}^{\mathrm{BHPT, amp}}(t)$ and $R_{\ell m}^{\mathrm{BHPT, freq}}$, for ppBHPT waveforms.
We also estimate the post-adiabatic (PA) orbital-phase correction due to spin,
\begin{equation}
\delta \phi_{\rm orb}^{\rm PA, spin}(t;\chi) = \phi_{\rm orb}^{\rm NR}(t;\chi) 
- \bigl[\phi_{\rm orb}^{\rm ppBHPT}(t;\chi)
+ \delta\phi_{\rm orb}^{\rm PA, no\_spin}(t)\bigr],
\label{eq:pa_est}
\end{equation}
with
\begin{equation}
\delta\phi_{\rm orb}^{\rm PA, no\_spin}(t) = \phi_{\rm orb}^{\rm NR}(t;\chi=0) - \phi_{\rm orb}^{\rm ppBHPT}(t;\chi=0)\,.
\end{equation}
Here $\delta\phi_{\rm orb}^{\rm PA, no\_spin}$ represents beyond-adiabatic corrections (all PA orders) in the non-spinning case (the orbital phase is extracted as $\phi_{\rm orb} = \frac{1}{2} \arg h_{22}$). Adding this to the adiabatic ppBHPT phase with spin yields a model that includes all non-spinning PA effects but only adiabatic spin effects. Comparing with $\phi_{\rm orb}^{\rm NR}(t;\chi)$ provides an estimate of spin-dependent corrections beyond the adiabatic approximation. Following Ref.~\cite{vandeMeent:2020xgc}, we compare dephasing to reference thresholds of $\pi/4$ and $\pi/16$.

\begin{figure}
\includegraphics[scale=0.57]{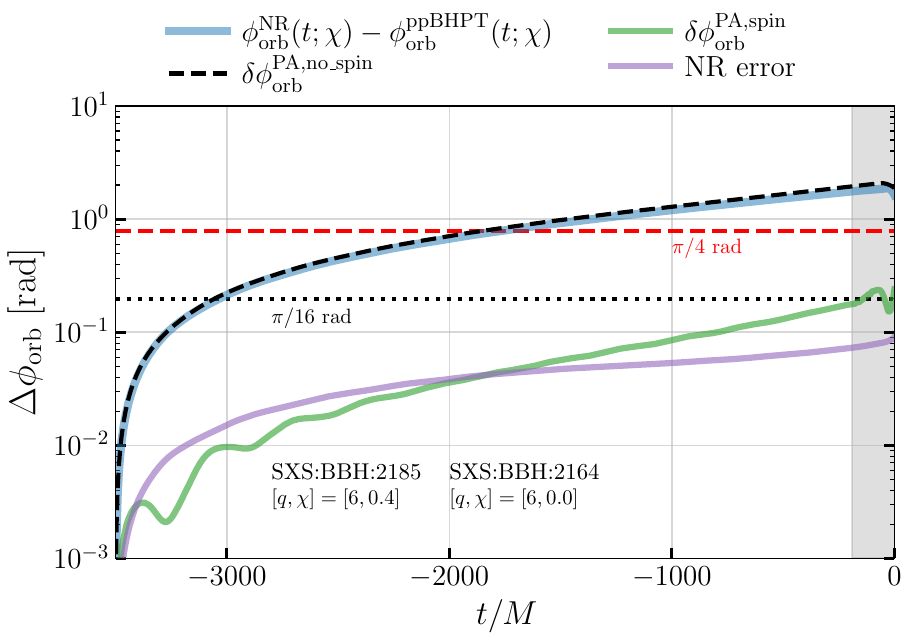}
\caption{Dephasing between NR (\texttt{SXS:BBH:2185}) and ppBHPT orbital phases for $[q,\chi]=[6,0.4]$. The dephasing (solid blue) grows significantly during the inspiral. Combining post-adiabatic phase information from the non-spinning case with the adiabatic ppBHPT spin phase reduces the dephasing by 1–2 orders of magnitude (solid green), demonstrating the effectiveness of adiabatic spin information. The residual $\delta \phi_{\rm orb}^{\rm PA,spin}$, defined in Eq.~\eqref{eq:pa_est}, captures remaining phase errors from neglecting higher-order spin-dependent PA terms. Horizontal lines indicate thresholds of $\pi/4$ (dashed red) and $\pi/16$ (dotted black). The shaded gray region marks the transition from inspiral to geodesic plunge. The dephasing between the two highest-resolution NR simulations is also shown as a benchmark (solid purple).}
\label{fig:delta_phase_q6_spin_0p4}
\end{figure}

\noindent\textbf{\textit{Comparison between NR and ppBHPT waveforms}: }
Figure~\ref{fig:R_nr_bhpt_example} shows the spin-enhancement quantities for two binaries across three modes, $[\ell, m] = [(2,2), (3,3), (4,4)]$, with parameters $[q, \chi] = [8, -0.5]$ and $[6, 0.4]$, capturing both mass-ratio and spin dependence (corresponding NR simulations are given in the supplemental material~\cite{SupplementalMaterial}). For both systems, the ppBHPT amplitude enhancement factors agree visually with NR across all modes. The frequency enhancement ratios are nearly identical across modes, so we show only the $(2,2)$ mode. While the frequency ratios remain close, their agreement is slightly weaker than for amplitudes. Overall, the consistency between NR and ppBHPT is quite good, given that ppBHPT uses only the lowest adiabatic order (0PA).
We next compute the spin-enhancement factor for binaries with $q=\{5,10,15\}$ and fixed spin $\chi=0.5$. Figure~\ref{fig:q_5_10_15} shows the resulting $(2,2)$ mode amplitude ratios, $R_{2,2}^{\mathrm{amp}}$, with corresponding ppBHPT predictions (corresponding NR simulations are given in the supplemental material~\cite{SupplementalMaterial}). The ppBHPT results closely track NR through the adiabatic inspiral, with with deviations becoming more pronounced once the ppBHPT systems enter the geodesic plunge (gray shaded regions).

\begin{figure}
\includegraphics[width=\columnwidth]{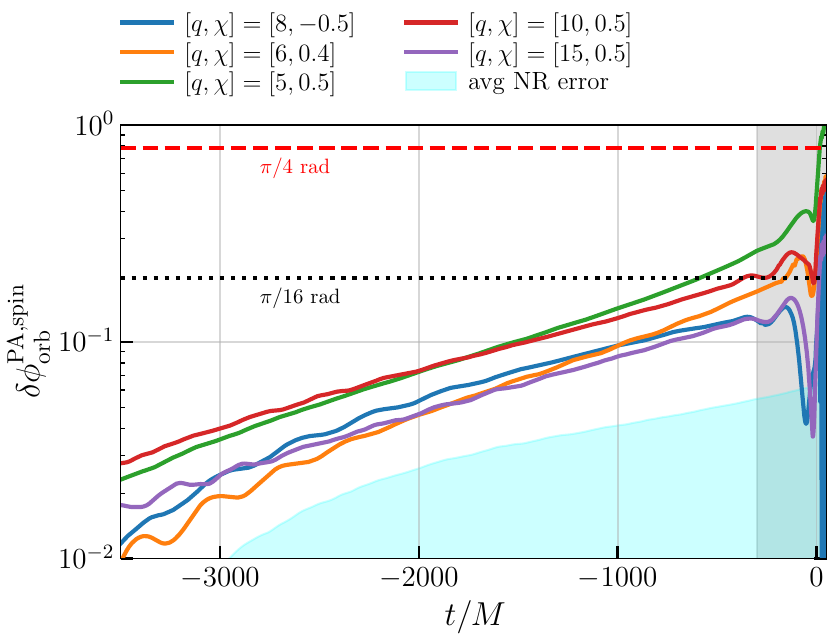}
\caption{We show the estimated phase correction that arises from neglecting higher-order spin-dependent PA terms, defined in Eq.(\ref{eq:pa_est}), for all BBHs considered in Figure~\ref{fig:R_nr_bhpt_example} and \ref{fig:q_5_10_15}. Horizontal lines provide reference thresholds of $\pi/4$ (dashed red) and $\pi/16$ (dotted black). Shaded gray region indicates the time span from the end of the adiabatic inspiral through the transition and into the geodesic plunge in the ppBHPT waveforms.
Additionally, we show the average dephasing between the two highest-resolution NR simulations for the five BBH systems (cyan-shaded region) as a benchmark.
}
\label{fig:delta_1pa_spin}
\end{figure}

In Figure~\ref{fig:delta_phase_q6_spin_0p4}, we show the dephasing between NR and ppBHPT waveforms (solid blue) for one binary from Figure~\ref{fig:R_nr_bhpt_example}, which exceeds 1 radian. Most of this arises from missing non-spinning PA corrections, $\delta\phi_{\rm orb}^{\rm PA,no\_spin}$ (black dashed). To quantify spin effects, we compute the estimated correction $\delta\phi_{\rm orb}^{\rm PA,spin}$ on top of the non-spinning phase. This estimated spin-induced correction remains below $\pi/16$ throughout the evolution. We show $\delta\phi_{\rm orb}^{\rm PA,spin}$ for all five binaries in Figures~\ref{fig:R_nr_bhpt_example} and \ref{fig:q_5_10_15}, computed using Eq.~(\ref{eq:pa_est}) (see Figure~\ref{fig:delta_1pa_spin}). In all cases, cumulative dephasing is mostly below $\pi/16$; near merger it may exceed this for $q \lesssim 5$ but remains well below $\pi/4$. This indicates that, once non-spinning PA corrections are included, additional spin-dependent corrections are likely small for the range of spins considered here.

\begin{figure}
\includegraphics[width=\columnwidth]{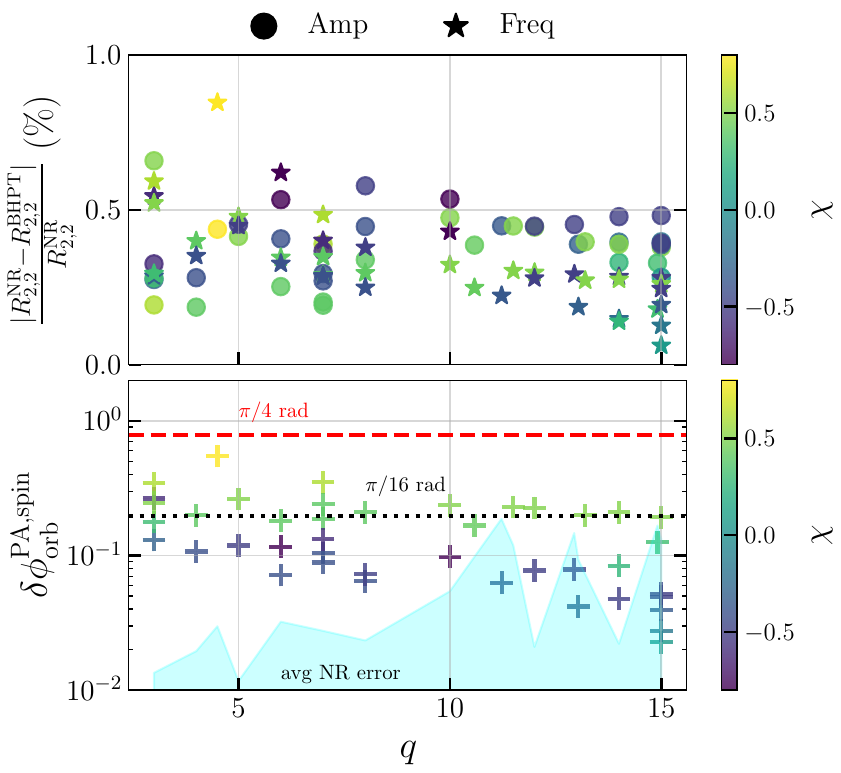}
\caption{We show the average percentage error in the $(2,2)$ mode amplitude and frequency spin-enhancement ratios from Eq.~\eqref{eq:sef} -- comparing spinning and non-spinning waveforms -- computed using NR and ppBHPT for a total of 43 NR simulations obtained from the SXS catalog (\textit{upper panel}). The corresponding estimated cumulative PA phase corrections associated with spin effects are shown in the lower panel. Additionally, we show the dephasing between the two highest-resolution NR simulations for each mass ratio (cyan-shaded region) as a benchmark.}
\label{fig:ratio_errors}
\end{figure}

We further quantify the agreement between NR and ppBHPT using 43 NR simulations from the SXS catalog with $q\geq3$ and $-0.8\leq\chi\leq0.8$, within the training domain of the \texttt{BHPTNRSur2dq1e3} surrogate~\cite{Rink:2024swg}, used here to generate ppBHPT waveforms (with NR calibration turned off so as to reproduce the Teukolsky solver waveforms). For each case, we also compute the cumulative dephasing due to PA spin corrections from the start of the waveform to $t=-275M$, marking the end of the adiabatic inspiral. Non-spinning reference waveforms are obtained from the \texttt{NRHybSur2dq15} surrogate~\cite{yoo2022targeted}, restricted to the final $5000M$ (or available NR length). Figure~\ref{fig:ratio_errors} shows that the average relative errors in the $(2,2)$ amplitude and frequency enhancement ratios (computed up to $t=0$) remain below $1\%$, with similar behavior in higher modes. For $\chi \leq 0.5$ and $q \gtrsim 8$, the dephasing from spin-dependent PA effects is below $\pi/16$, indicating a relatively small effect. As expected, both enhancement-ratio differences and dephasing decrease with increasing mass ratio.

\begin{figure}
\includegraphics[width=\columnwidth]{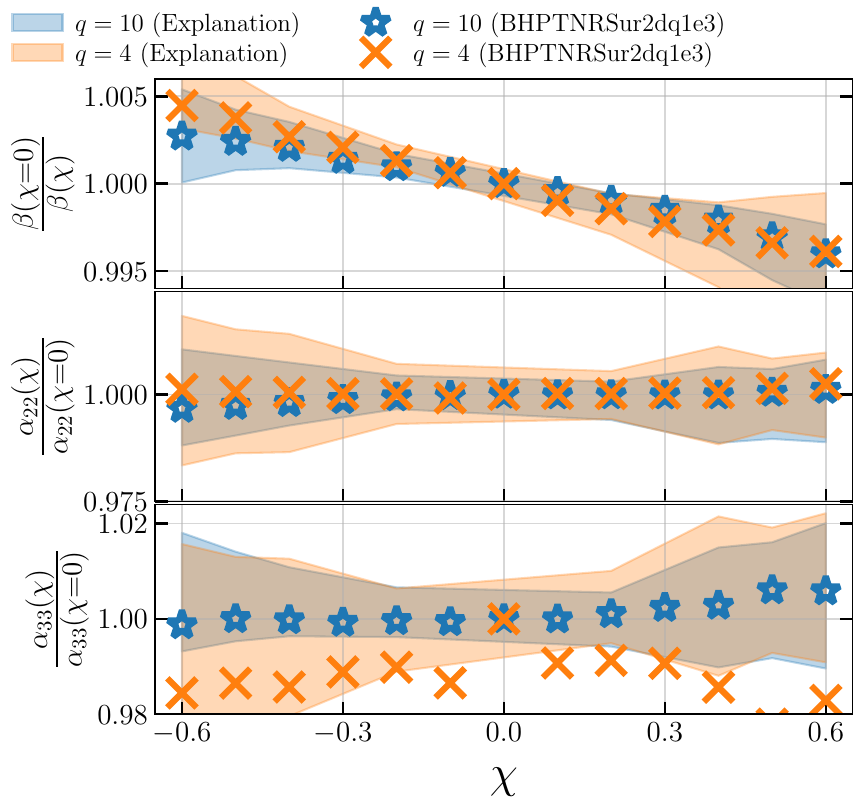}
\caption{We show the ratio of the model parameters $\beta$ (upper panel), $\alpha_{22}$ (middle panel), and $\alpha_{33}$ (lower panel)—used in the construction of the \texttt{BHPTNRSur2dq1e3} waveform model—between a spinning binary and its non-spinning counterpart, for two mass-ratio values in the comparable-mass regime: $q = 4$ (blue stars) and $q = 10$ (orange crosses). For comparison, we also show ranges of the ratios (labeled ``Explanation") $R_{22}^{\mathrm{BHPT, freq}} / R_{22}^{\mathrm{NR, freq}}$, $R_{22}^{\mathrm{NR, amp}} / R_{22}^{\mathrm{BHPT, amp}}$, and $R_{33}^{\mathrm{NR, amp}} / R_{33}^{\mathrm{BHPT, amp}}$, computed at 25 equally spaced time samples. The connection between these ratios and the calibration parameters is explained in Eq.~\eqref{eq:relations2}.}
\label{fig:alpha_beta_explained}
\end{figure}

\begin{figure*}
\includegraphics[scale=0.55]{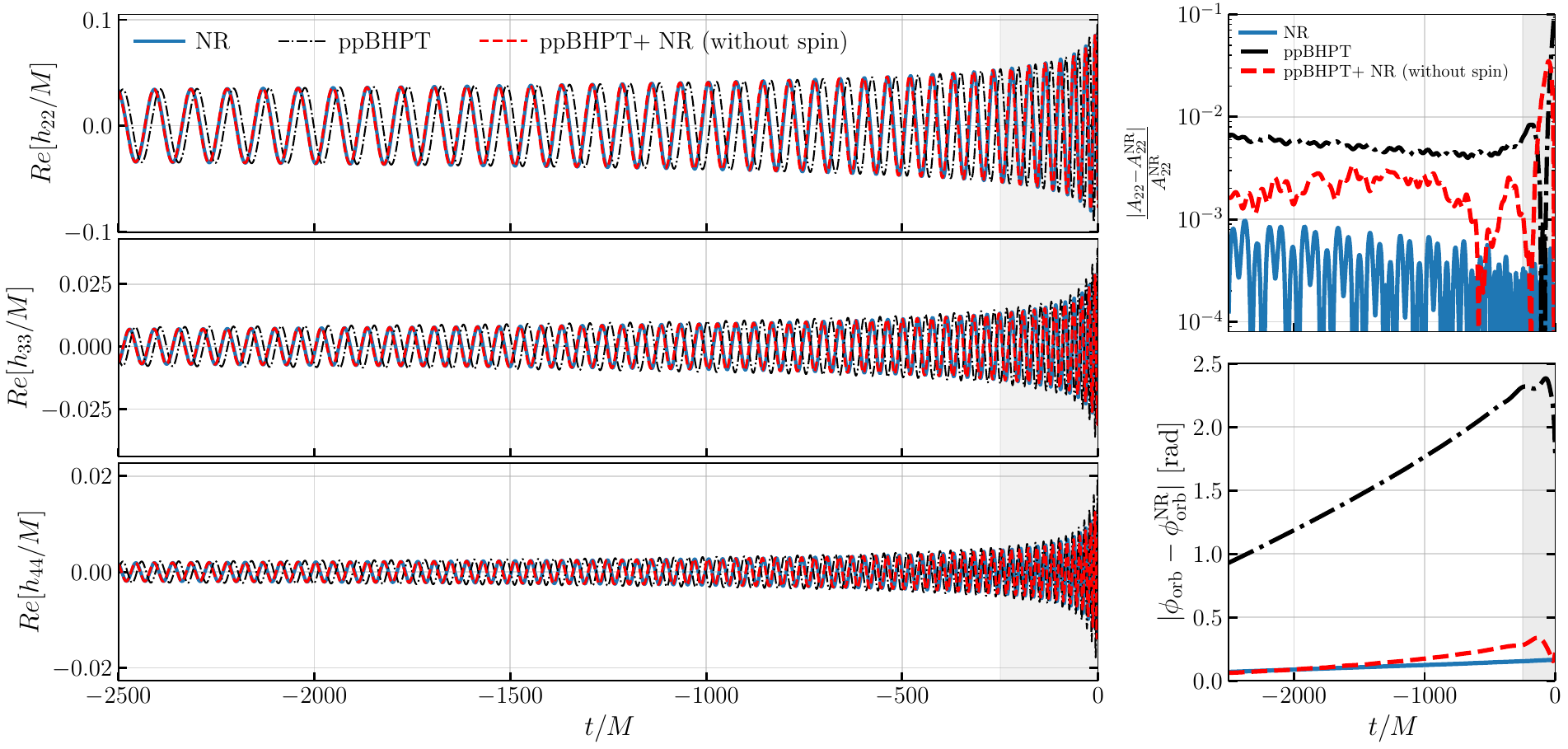}
\caption{{\bf Left}: We show the final $2500M$ of the $(2,2)$, $(3,3)$, and $(4,4)$ modes for the NR simulation \texttt{SXS:BBH:2464} (blue solid lines; until merger), characterized by $[q, \chi] = [15, 0.5]$, alongside the corresponding ppBHPT predictions (black dashed lines) after a time and phase shifts. While ppBHPT alone does not fully match NR, combining the non-spinning NR prediction (\texttt{SXS:BBH:2477}) with spin effects extracted from adiabatic ppBHPT waveforms yields good agreement, with deviations becoming more pronounced once the ppBHPT systems enter the geodesic plunge (gray shaded regions). 
{\bf Right}: Corresponding $(2,2)$ amplitude and orbital phase errors relative to NR, along with errors between the two highest-resolution NR simulations as a benchmark.}
\label{fig:modelling_implication}
\end{figure*}

\noindent\textbf{\textit{Insights into the BHPTNRSurrogate(s) modelling approach}: }
We now connect our results to the \texttt{BHPTNRSurrogate(s)} family of waveform models~\cite{Rink:2024swg,Islam:2022laz,Rifat:2019ltp}. These models introduce a simple rescaling of ppBHPT waveforms,
$h_{\ell m}^{\mathrm{NR}}(t_{\mathrm{NR}}) \sim \alpha_{\ell}\, h_{\ell m}^{\mathrm{BHPT}}\left( \beta\, t_{\mathrm{BHPT}} \right)$,
where $h_{\ell m}^{\mathrm{NR}}$ and $h_{\ell m}^{\mathrm{BHPT}}$ denote NR and ppBHPT waveforms as functions of $t_{\mathrm{NR}}$ and $t_{\mathrm{BHPT}}$. The calibration parameters $\alpha_{\ell}$ and $\beta$ are obtained by fitting ppBHPT waveforms to NR. Comparing non-spinning~\cite{Islam:2022laz} and spinning~\cite{Rink:2024swg} calibrations shows that $\alpha_\ell$ is nearly independent of spin, while $\beta$ has a weak linear spin dependence. We interpret this behavior using the spin-enhancement ratios $R_{\ell m}^{\mathrm{amp}}$ and $R_{\ell m}^{\mathrm{freq}}$.
Starting from~\cite{Islam:2023jak},
\begin{equation}
A_{\ell m}^{\mathrm{NR}}(t_{\mathrm{NR}}) \approx \alpha_{\ell}\, A_{\ell m}^{\mathrm{BHPT}}\left( \beta\, t_{\mathrm{BHPT}} \right); 
\quad
\omega_{\mathrm{orb, NR}} \approx \frac{1}{\beta}\, \omega_{\mathrm{orb, BHPT}} \,,
\label{eq:freq_rescale}
\end{equation}
we obtain
\begin{equation} \label{eq:relations2}
\frac{R_{\ell m}^{\mathrm{NR, amp}}(q,\chi)}{R_{\ell m}^{\mathrm{BHPT, amp}}(q,\chi)} \approx \frac{\alpha_\ell(q,\chi)}{\alpha_\ell(q,\chi = 0)},
\quad
\frac{R_{\ell m}^{\mathrm{BHPT, freq}}(q,\chi)}{R_{\ell m}^{\mathrm{NR, freq}}(q,\chi)} \approx \frac{\beta(q,\chi)}{\beta(q,\chi = 0)}.
\end{equation}
The left-hand sides are time-dependent, while the right-hand sides are constants. Since the time variation is weak, we evaluate the left-hand sides at 25 uniformly spaced points from the start of the NR waveform to near merger and take the average, with the spread indicating temporal variation. Figure~\ref{fig:alpha_beta_explained} shows these comparisons for $q=4$ and $q=10$, with shaded regions indicating variation in the enhancement ratios. We also show the ratios of model parameters between spinning and non-spinning binaries for $\beta$ (top), $\alpha_{22}$ (middle), and $\alpha_{33}$ (bottom), used in the \texttt{BHPTNRSur2dq1e3} model. The observed spin dependence of $\alpha$ and $\beta$ is explained by the amplitude and frequency relations in Eq.~\eqref{eq:relations2}.

\noindent\textbf{\textit{A possible new approach to waveform modelling}: }
We now consider a waveform modeling strategy in which the non-spinning sector is informed by NR data, while spin effects are taken from adiabatic ppBHPT waveforms. This approach may be particularly useful when NR simulations -- especially at intermediate mass ratios and high spins, which can take months or longer -- are significantly more expensive than ppBHPT simulations. In such cases, ppBHPT waveforms could assist in building accurate waveform models while requiring fewer NR training simulations.
In Figure~\ref{fig:modelling_implication}, we show the $(2,2)$, $(3,3)$, and $(4,4)$ modes for the NR simulation \texttt{SXS:BBH:2464}, with $[q,\chi]=[15,0.5]$, alongside ppBHPT predictions and hybrid model results (including amplitude and phase errors) obtained by combining non-spinning NR data with ppBHPT spin effects. While ppBHPT alone is insufficient to reproduce the NR data across all modes accurately, the hybrid waveform agrees well with the NR waveform across all modes. This hybrid waveform should be viewed as a proof-of-concept illustrating that simple NR–perturbation-theory combinations can achieve close agreement; further development is needed before direct use in data analysis. The calibration techniques of Ref.~\cite{Rink:2024swg} provide one possible path forward.

\noindent\textbf{\textit{Concluding remarks}: }
This work shows that ppBHPT, even at low order in mass ratio, captures some effects of the primary's spin in the waveform with mass ratios as low as $q \approx 5$ and spins $\chi \lesssim 0.5$, with better accuracy than might be expected from its nominal regime of validity. In particular, we find that spin effects in ppBHPT waveforms (without spin information beyond adiabatic order) remain in good agreement with NR waveforms over the final $\sim20$ orbits, and in some cases provide a more faithful description than commonly used post-Newtonian models (further details can be found in the Supplementary Material). These results support the incorporation of information computed within ppBHPT into existing waveform modelling efforts, and suggest that such an approach may be particularly beneficial for intermediate mass ratio systems where NR simulations can be prohibitively expensive across much of the parameter space. Our work also suggests that while extending post-adiabatic calculations~\cite{Pound:2021qin,Miller:2020bft,Wardell:2021fyy} to Kerr black holes will be technically challenging, modeling their effects across parameter space may be comparatively straightforward (as in Ref.~\cite{Rink:2024swg}). For spins $\chi \lesssim 0.5$, post-adiabatic differences between spinning and non-spinning binaries remain small (especially in the waveform amplitudes) over typical NR durations of $\sim 4000M$ (about 20 orbital cycles), suggesting that Schwarzschild second-order results combined with first-order spin corrections could form the basis of accurate spin-dependent models. Future work in this direction should include spin-precession and eccentricity effects.


\begin{acknowledgments}
We thank Ritesh Bacchar, Nils Deppe, Achamveedu Gopakumar, Rahul Kashyap, Ajit Kumar Mehta, Michael Pürrer, Tejaswi Venumadhav, Harald Pfeiffer, Maarten van de Meent, and Barry Wardell for helpful discussions. 
This research was supported in part by the National Science Foundation under Grant No. NSF PHY-2309135 and the Gordon and Betty Moore Foundation Grant No. GBMF7392.

Use was made of computational facilities purchased with funds from the National Science Foundation (CNS-1725797) and administered by the Center for Scientific Computing (CSC). The CSC is supported by the California NanoSystems Institute and the Materials Research Science and Engineering Center (MRSEC; NSF DMR 2308708) at UC Santa Barbara. G.K. acknowledges support from NSF Grants No. PHY-2307236 and DMS-2309609. 
G.K. acknowledges support from NSF Grants No. PHY-2307236 and DMS-2309609.
S.F. acknowledges support from NSF Grants No. AST-2407454 and PHY-2110496.
This work was partly supported by UMass Dartmouth's Marine and Undersea Technology (MUST) research program funded by the Office of Naval Research (ONR) under grant no. N00014-23-1-2141.
Some computations were performed on the UMass-URI UNITY HPC/AI cluster at the Massachusetts Green High-Performance Computing Center (MGHPCC).
\end{acknowledgments}

\bibliography{references}

\appendix 
\section*{Supplemental Material}

\noindent\textbf{\textit{NR Simulations used in the main text}: }
For Fig.~1 in the main text, the corresponding NR simulations from the SXS catalog are \texttt{SXS:BBH:3623} and \texttt{SXS:BBH:2185}, with associated non-spinning simulations \texttt{SXS:BBH:3622} ($q=8$) and \texttt{SXS:BBH:2164} ($q=6$).
For Fig.~2, the spinning NR simulations are \texttt{SXS:BBH:2329}, \texttt{SXS:BBH:2485}, \texttt{SXS:BBH:2469}, and \texttt{SXS:BBH:2464}, with corresponding non-spinning simulations \texttt{SXS:BBH:2325}, \texttt{SXS:BBH:2387}, \texttt{SXS:BBH:0185}, and \texttt{SXS:BBH:2477}.

\noindent\textbf{\textit{Comparison against PN approximations}: }
Another framework for generating waveforms is the post-Newtonian (PN) approximation (for a detailed review, see Ref.~\cite{Blanchet:2013haa}), in which the Einstein equations are expanded in powers of small velocity, $\frac{v}{c}$ (where $v$ is the velocity of the black holes and $c$ is the speed of light), and weak gravitational fields. The accuracy of PN approximation depends on the order to which the expansion is carried out~\cite{Arun:2004hn,Blanchet:2004bb,Blanchet:2013haa,Kidder:1992fr,Paul:2022xfy,Blanchet:2006gy,Porto:2010zg,Bohe:2013cla,Marsat:2013caa,Cho:2022syn}. While PN can accommodate arbitrary mass ratios, it becomes increasingly inaccurate in regimes of high velocity or strong gravitational fields. Therefore, it cannot provide accurate waveforms in the late inspiral or merger stages of the evolution~\cite{Boyle:2007ft,Hannam:2007wf,Pan:2007nw,Hannam:2007ik}. 

\begin{figure}
\includegraphics[width=\columnwidth]{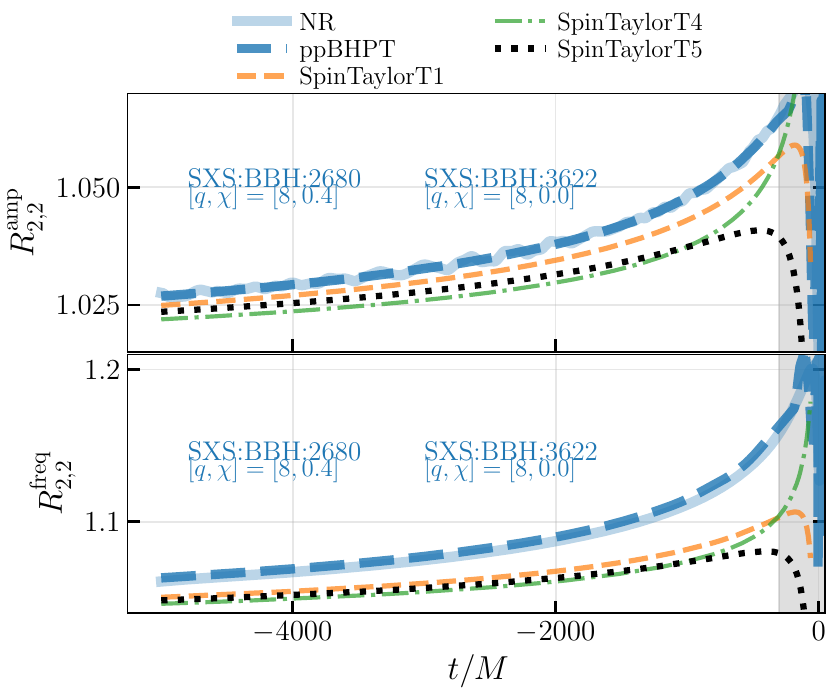}
\caption{We show the ratios of $(2,2)$ mode spin enhancement ratios for the amplitudes and frequencies for $[q, \chi] = [8, 0.4]$. Solid blue lines represent NR values obtained from the SXS data, while dashed blue lines indicate the corresponding ppBHPT results, rescaled to the same mass scale. In addition, we include post-Newtonian (PN) results computed using the \texttt{SpinTaylorT1}, \texttt{SpinTaylorT4} and \texttt{SpinTaylorT5} approximants. The latter incorporates five additional pseudo-PN terms in both amplitude and frequency, with coefficients calibrated to NR data as described in Ref.~\cite{Estelles:2020osj}. Shaded gray regions indicate the time span from the end of the adiabatic inspiral through the transition and into the geodesic plunge in the ppBHPT waveforms.
}
\label{fig:R_nr_bhpt_phenom_example}
\end{figure}

We compute the same set of spin-enhancement factors (defined in Eq.(3) in the main text) using PN approximations, specifically the \texttt{SpinTaylorT1}, \texttt{SpinTaylorT4}~\cite{Buonanno:2009zt} and \texttt{SpinTaylorT5}~\cite{Damour:2001bu} approximants\footnote{We access these approximants from the \texttt{LALSuite} package~\cite{lalsuite}.}.
These approximants differ in how the waveforms are expanded. All are truncated at 3.5PN order and include nonlinear mass-ratio ($q$) effects. Spin effects are included up to 3.5PN order (including $q$--$\chi$ coupling terms), though partial cubic-in-spin terms are not present~\cite{Buonanno:2009zt,Damour:2001bu}. 
Figure~\ref{fig:R_nr_bhpt_phenom_example} shows the PN enhancement factors for the $(2,2)$ mode amplitude and frequency for one particular case, along with the corresponding NR and ppBHPT values. 
We find that neither the amplitude enhancement factor nor the frequency enhancement factor from any of the three PN waveform approximants match the NR values at any point during the binary evolution. Among the three, the \texttt{SpinTaylorT1} approximant shows the closest agreement with NR for both amplitude and frequency enhancement factors. However, all PN approximants exhibit significant deviations from NR as the binary approaches merger, with their predictions breaking down and diverging rapidly near this stage. While typical waveform diagnostics focus on dephasing between models, to better connect with the main-body of the paper, we choose to work with the amplitude and frequency enhancement factors. However, given the differences seen in Fig.~\ref{fig:R_nr_bhpt_phenom_example}, the dephasing between the NR and the PN variants will be around $\sim 10$ radians.

\end{document}